\documentclass[conference]{IEEEtran}
\IEEEoverridecommandlockouts
\usepackage{cite}
\usepackage{amsmath,amssymb,amsfonts}
\usepackage{algorithmic}
\usepackage{graphicx, comment}
\usepackage{textcomp, enumitem}
\usepackage{xcolor}
\usepackage[a4paper, total={184mm,239mm}]{geometry}
\def\BibTeX{{\rm B\kern-.05em{\sc i\kern-.025em b}\kern-.08em
    T\kern-.1667em\lower.7ex\hbox{E}\kern-.125emX}}

\usepackage{diagbox}
\usepackage{multirow}
\newcommand\TODO[1]{\textcolor{red}{[TODO]: #1}}

\usepackage{fancyhdr}
\begin{document}

\title{BitPrefill: W1.58A4 Tensor-Core Acceleration for Compute-Bound LLM Prefill
}

\author{
  \IEEEauthorblockN{
  Fang Chen\IEEEauthorrefmark{1}, 
  Xuming Chen\IEEEauthorrefmark{1},
  Sanmukh Kuppannagari\IEEEauthorrefmark{1},
  Gourav Datta\IEEEauthorrefmark{1}
  }
  \IEEEauthorblockA{\IEEEauthorrefmark{1}\textit{Case Western Reserve University}, Cleveland, US \\
  }
}

\maketitle

\begin{abstract}
Low-bit quantization reduces large language model (LLM) inference
cost, yet implementations optimized for memory-bound decode can
leave compute-bound prefill on BF16 tensor cores. This gap matters
for input-heavy workloads, including long-context and multimodal
applications, where prefill can account for a substantial fraction
of inference latency. We propose a W1.58A4 operator that quantizes weights to ternary values and activations to 4-bit, while executing the linear projection on native INT4 general matrix multiply (GEMM) tensor cores, turning prefill from a full-precision matrix multiply into a low-bit one. We deploy the operator for BitNet-b1.58 on a single RTX 4090 and a Jetson Orin Nano edge device, and report GEMM-kernel latency, end-to-end LLMs prefill latency, and task accuracy against bf16 and BitNet/BitBLAS baselines. Our proposed kernel achieves ${\sim}1.42{\times}$ acceleration compared to the original BF16 prefill phase, with negligible reduction on model accuracy.
\end{abstract}

\section{Introduction}

Efficient large language model (LLM) inference on constrained hardware
has become increasingly important as LLMs move from data-center servers
to consumer GPUs and edge platforms. Low-bit quantization is one of the
primary techniques for reducing inference cost. Existing
systems~\cite{bitnetb1_58-2b4t, bitnet_cpp, bitnet1_58, onebit, pbllm, billm}
commonly quantize weights to 4 bits or below while keeping activations
in BF16 or INT8, and execute the resulting linear layers using
mixed-precision GEMM kernels. This design is highly effective for the
\emph{decode} stage, where the model generates one token at a time. In
this regime, the GEMM has a very small token dimension
($M=1$ per sequence), leading to low arithmetic intensity and
memory-bound execution. Weight movement therefore becomes a dominant
cost, so reducing weight precision directly lowers memory traffic and
can substantially accelerate decode. However, the arithmetic itself
typically still executes at a higher precision, such as INT8 or BF16
depending on the activation format.

The same approach is less effective for \emph{prefill}, which is the
gap targeted in this work. Prefill processes the entire input context
before generation begins, producing GEMMs with a much larger token
dimension $M$ and higher arithmetic intensity. As $M$ grows, execution
becomes increasingly compute-bound, so reducing weight traffic alone
provides diminishing returns. Thus, accelerating prefill requires
lowering the arithmetic precision of the GEMM itself by reducing both
weight and activation precision, rather than only compressing the
stored weights.

This distinction is increasingly relevant for workloads with long input
contexts relative to their outputs, including summarization~\cite{zhang2025systematic, zhang2024benchmarking}, long-context
reasoning~\cite{bai2025longbench, ling2025longreason}, and multimodal inference~\cite{wu2023next, zhan2024anygpt, islam2025gpt}. For example, a $336\times336$ image
with $16\times16$ patches produces 441 visual patch tokens before any
model-specific compression, so multiple frames can quickly enlarge the
prefill sequence. In such input-dominated workloads, prefill can become
a major contributor to time-to-first-token and end-to-end latency.

\begin{figure}
    \centering \includegraphics[width=0.9\linewidth]{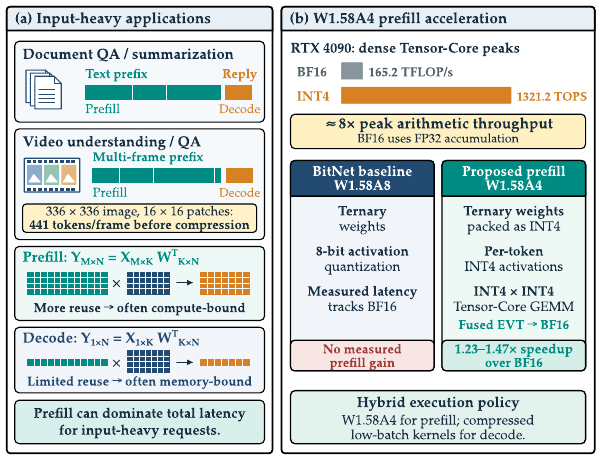}
    \vspace{-4mm}
    \caption{\textbf{Motivation for our proposed method}. Input-heavy applications make prefill compute-bound, while batch-1 decode remains memory-bound. W1.58A4 exploits INT4 Tensor Cores for faster prefill while retaining compressed kernels for decode.}
    \label{fig:sigprocess-i}
    \vspace{-3mm}
\end{figure}

This limitation is particularly important for ternary LLMs such as
BitNet~\cite{bitnet1_58, bitnet_cpp}. A ternary weight takes one of three values,
$\{-1,0,+1\}$, corresponding to an information-theoretic precision of
$\log_2(3)\approx1.58$ bits; hence the notation \emph{W1.58}. Ternary
models are attractive for edge inference because they substantially
reduce model storage and weight-transfer cost. However, a W1.58A16
implementation still performs its dominant prefill computation on BF16
tensor cores because the activation operand remains BF16. Thus, it
reduces memory traffic while leaving the compute roofline largely
unchanged. On an RTX 4090, the theoretical peak throughput is roughly
1321 INT4 TOPS versus 165 BF16 TFLOPS, exposing an approximately
$8\times$ gap in available arithmetic throughput.

We therefore develop a \textbf{W1.58A4} execution path in which BitNet
weights remain algorithmically ternary but are represented using signed
INT4 fields for hardware execution, while activations are quantized to
INT4. This enables the prefill GEMM to execute on native INT4 tensor-core
MMA instructions available on Ampere and Ada GPUs. W1.58 describes the
algorithmic weight precision, whereas INT4 is the physical execution
format required by current commodity tensor cores. We implement the
operator using CUTLASS and evaluate it on both an RTX 4090 GPU and a
Jetson Orin Nano edge device.

Achieving a fast INT4 GEMM alone is not sufficient for end-to-end
acceleration. First, INT4 GEMM accumulates into INT32, and a separate
dequantization kernel would require additional passes over the
$M\times N$ output. We therefore fuse per-channel weight scaling and
per-token activation scaling into the GEMM epilogue using an Epilogue
Visitor Tree (EVT), directly emitting BF16 outputs. Second, the optimal
INT4 tile configuration depends strongly on the prefill token dimension
$M$. We therefore precompile a set of candidate tile configurations,
select and cache the best configuration for each problem shape, and
combine this with Stream-K~\cite{streamk} scheduling to improve utilization for
irregular tile counts.

In summary, this paper makes three contributions:
\begin{itemize}[leftmargin=*]

\item We develop a \textbf{W1.58A4 INT4 Tensor-Core operator} for
compute-bound LLM prefill, enabling ternary BitNet weights and
4-bit activations to execute through native INT4 GEMM on both RTX 4090
and Jetson Orin Nano platforms.

\item We introduce a \textbf{shape-adaptive kernel selection (SAKS)}
strategy that selects among precompiled GEMM candidates for each
$(M,N,K)$ shape and device, including tile geometry, pipeline depth,
and optional Stream-K work decomposition to improve utilization across
prefill lengths.

\item We propose an \textbf{EVT-fused dequantization GEMM} that applies
per-channel weight scales and per-token activation scales directly in
the epilogue, converting INT32 accumulators to BF16 outputs without
materializing an intermediate INT32 tensor or launching a separate
dequantization kernel.

\end{itemize}
\section{Related Work}

\subsection{Low-bit Large Language Models}
BitNet~\cite{bitnet} introduced end-to-end 1-bit transformer training through a quantization-aware BitLinear layer, while BitNet-b1.58~\cite{bitnet1_58} extended this to ternary weights with absmean scaling and near full-precision perplexity. BitNet-b1.58-2B4T~\cite{bitnetb1_58-2b4t} is the largest open model in this family. Spectra~\cite{spectra} and ParetoQ~\cite{paretoq} further show strong size--accuracy tradeoffs for ternary models, whereas PB-LLM~\cite{pbllm}, BiLLM~\cite{billm}, and OneBit~\cite{onebit} obtain 1-bit weights from pretrained models. However, these methods retain 8--16-bit activations, leaving GEMM execution at INT8/BF16 precision. BitNet a4.8~\cite{bitnet_a4_8} introduces 4-bit activations through hybrid quantization and sparsification, but does not provide a released GPU kernel. Our work provides the required INT4 tensor-core operator for W1.58A4 execution and evaluates it end-to-end on consumer GPU and edge platforms.

On the systems side, bitnet.cpp~\cite{bitnet_cpp} is the official CPU engine for ternary LLMs and builds on T-MAC~\cite{tmac}, which uses table lookups over precomputed low-bit partial sums to avoid explicit dequantization. LUT Tensor Core~\cite{luttensorcore} extends this approach through hardware/software co-design. These works primarily target memory-bound decoding on CPUs or custom hardware, whereas we target compute-bound prefill on commodity GPU tensor cores.

\subsection{GEMM Kernels for Low-bit LLMs}
 CUTLASS~\cite{cutlass} exposes native INT4 MMA on Ampere/Ada GPUs through templated threadblock, warp, and instruction tiles, multi-stage \texttt{cp.async} pipelines, and runtime selection among pre-instantiated kernels.
 Stream-K~\cite{streamk} splits a GEMM into mainloop iterations shared equally among persistent threadblocks, with partial tiles reconciled through a workspace, so utilization no longer depends on how the tile count divides the SM count.
 It balances how work is distributed but not how efficient each tile is.
 BitBLAS~\cite{bitblas} instead uses TVM-based compilation and per-shape JIT autotuning for mixed-precision GEMMs such as FP16$\times$INT4 and INT4$\times$INT4. 
 The GEMM epilogue processes accumulator fragments after the
matrix reduction and before storing the output. 
The Epilogue Visitor Tree (EVT)~\cite{evt} represents GEMM
post-processing as a compile-time expression tree executed over
accumulator fragments within the epilogue. It supports composable
broadcasts, type conversions, reductions, and element-wise operations,
allowing post-processing that would otherwise require separate kernels
to be fused into GEMM. We use EVT to apply per-token and per-channel
scales to INT32 accumulators and directly emit BF16 outputs.

\section{Proposed W1.58A4 Prefill Execution}
\label{sec:method}

\subsection{System Overview and Hybrid Execution}

A BitNet linear layer computes $Y=XW^{T}$, where
$X\in\mathbb{R}^{M\times K}$ contains token representations,
$W\in\mathbb{R}^{N\times K}$ contains weights, and
$Y\in\mathbb{R}^{M\times N}$. Here, $M$ is the GEMM token dimension,
including the batch dimension when multiple sequences are flattened.
Prefill processes many tokens concurrently, producing a matrix--matrix
multiply with high arithmetic intensity, whereas autoregressive decode
typically has $M=1$ or a small effective batch and is dominated by
weight movement. We therefore use a hybrid policy: prefill uses the
proposed W1.58A4 Tensor-Core GEMM, while decode retains the
weight-compressed GEMV or low-batch kernel from the original BitNet
runtime.

\begin{figure}
    \centering \includegraphics[width=1\linewidth]{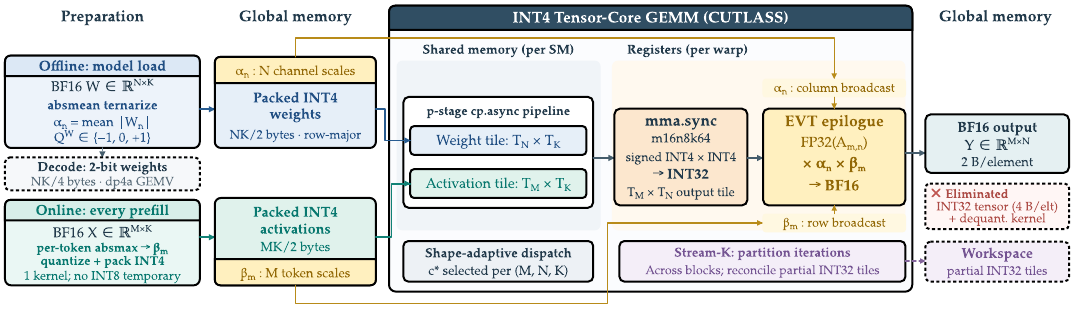}
    \vspace{-8mm}
    \caption{Prefill dataflow through the memory hierarchy. Packed INT4 operands feed Tensor-Core GEMM, while the fused EVT epilogue directly produces scaled BF16 outputs, eliminating the intermediate INT32 output tensor and separate dequantization kernel.}
    \label{fig:dataflow}
    \vspace{-3mm}
\end{figure}

Figure~\ref{fig:dataflow} shows the resulting prefill dataflow. Weights
are ternarized, packed, and transformed into the kernel layout once at
model load. During prefill, BF16 activations are quantized per token
and packed into signed INT4 fields. The packed operands are consumed by
an INT4 Tensor-Core GEMM with INT32 accumulation, and an EVT epilogue
applies the activation and weight scales before directly emitting BF16
outputs. This avoids materializing an intermediate INT32 output tensor
and launching a separate dequantization kernel.

\subsection{Arithmetic-Intensity Motivation}
\label{sec:arithmetic-intensity motivation}

Assuming each operand is read once and the output written once,
the idealized GEMM arithmetic intensity is
\begin{equation}
I(M,N,K)=\frac{2MNK}{b_aMK+b_wNK+b_oMN},
\label{eq:arithmetic_intensity}
\end{equation}
where $b_a$, $b_w$, and $b_o$ are the physical bytes per activation,
weight, and output element. Our packed INT4 GEMM with BF16 output
uses $b_a=b_w=0.5$ and $b_o=2$; this estimate excludes online
quantization, scale loads, and scheduling workspace traffic. When weight traffic dominates at $M=1$, $I\approx 2/b_w$.
Larger $M$ amortizes weight reads, approaching the compute-bound
regime when $I\geq\Pi_{\mathrm{peak}}/B_{\mathrm{mem}}$, where
$\Pi_{\mathrm{peak}}$ and $B_{\mathrm{mem}}$ denote peak dense
throughput at the chosen arithmetic precision and memory bandwidth.
W1.58A4 therefore targets prefill by reducing operand traffic and
enabling native INT4 computation. Realized gains also depend on
quantization overhead, occupancy, and scheduling efficiency.

\vspace{-1mm}
\subsection{Ternary-to-INT4 Mapping and Activation Quantization}
\label{sec:quantization}

\noindent\textbf{Ternary weights}: We use a per-output-channel absmean scale
\begin{equation}
 \alpha_n =
 \frac{1}{K}\sum_{k=1}^{K}|W_{n,k}|,\ 
 Q^W_{n,k} =
 \operatorname{clip}
 \left(
 \operatorname{round}\left(\frac{W_{n,k}}{\alpha_n}\right),
 -1,1
 \right).
 \label{eq:weight_quantization}
\end{equation}
The reconstructed weight is
$W_{n,k}\approx\alpha_nQ^W_{n,k}$, where
$Q^W_{n,k}\in\{-1,0,+1\}$.  
We store each ternary value in a signed INT4 field without
changing its numerical value. W1.58 denotes the three-valued
weight alphabet, whereas the execution representation occupies
four bits per weight. This mapping enables the quantized
activations and ternary weights to use the supported signed
INT4$\times$INT4 Tensor Core instructions. It trades the storage
density of a compact ternary encoding for direct execution on
commodity Tensor Cores.

\noindent\textbf{Per-token activation quantization}: For token $m$, we determine its dynamic range as
\begin{equation}
 a_m =
 \max\left(\max_k |X_{m,k}|,\epsilon\right),
 \qquad
 \beta_m = \frac{a_m}{7}.
 \label{eq:activation_scale}
\end{equation}
Here, $\epsilon>0$ prevents a zero activation scale. We then quantize the activation to the symmetric INT4 range:
\begin{equation}
 Q^X_{m,k} =
 \operatorname{clip}
 \left(
 \operatorname{round}\left(\frac{X_{m,k}}{\beta_m}\right),
 -7,7
 \right).
 \label{eq:activation_quantization}
\end{equation}
Thus, $X_{m,k}\approx\beta_mQ^X_{m,k}$. Per-token scaling adapts to
input-dependent activation ranges without requiring a calibration
dataset. The quantization kernel computes the rowwise maximum, derives
$\beta_m$, quantizes the row, and writes packed INT4 values in one
kernel invocation so that the intermediate unpacked integer tensor is
never materialized.

\noindent\textbf{Integer accumulation}: The Tensor Core mainloop computes $A_{m,n} =
 \sum_{k=1}^{K}Q^X_{m,k}Q^W_{n,k}$, using signed INT4 activations and ternary weights stored in signed INT4
fields. Since $Q^X_{m,k}\in\{-7,\ldots,7\}$ and
$Q^W_{n,k}\in\{-1,0,+1\}$, each product has magnitude at most $7$,
and hence $|A_{m,n}|\leq 7K$. For typical LLM dimensions
($K$ on the order of $10^3$--$10^4$), this bound remains several
orders of magnitude below the INT32 limit. The dequantized output is
\begin{equation}
Y_{m,n}\approx\beta_m\alpha_nA_{m,n}.
\label{eq:dequantization}
\end{equation}

\subsection{Packed Storage and Tensor-Core Dataflow}
\label{sec:hardware_mapping}

Two signed INT4 values are packed into each byte. Within a
logical row, elements $k=2j$ and $k=2j+1$ occupy the lower and
upper nibbles of byte $j$, respectively. Four-bit two's-complement
encoding represents $-1$ as \texttt{0xF}. For even $K$, the packed
activation and weight payloads occupy $MK/2$ and $NK/2$ bytes,
excluding scales and any alignment padding. 

Let $T_K$ denote the reduction depth processed in one mainloop
iteration. A full $T_M\times T_N$ output tile consumes a
$T_M\times T_K$ activation tile and a $T_N\times T_K$ weight
tile. Each activation is reused across $T_N$ output channels,
while each weight is reused across $T_M$ tokens. The packed
operand payload for this iteration is
$B_{\mathrm{stage}}=T_K(T_M+T_N)/2$ bytes, giving
\begin{equation}
I_{\mathrm{tile}}=
\frac{2T_MT_NT_K}{B_{\mathrm{stage}}}
=
\frac{4T_MT_N}{T_M+T_N}
\quad\mathrm{ops/byte}.
\label{eq:tile_reuse}
\end{equation}
This quantity describes computation per byte of staged operands,
excluding output traffic and repeated shared-memory fragment
accesses. Increasing $T_M$ or $T_N$ improves operand reuse,
whereas increasing $T_K$ alone does not change this ratio.
Instead, $T_K$ controls the operand-buffer size and the number
of mainloop iterations. These distinct effects motivate
considering output tiling and reduction depth separately.

Weights are quantized and packed during model initialization;
the packed weights and channel scales are reused across requests.
Activations are quantized and packed online using input-dependent
token scales. Vectorized global-memory accesses transfer the
packed operands into CUTLASS shared-memory layouts that support
warp-level operand-fragment loading. The mainloop operates on $T_M\times T_N$ output tiles using a
$p$-stage shared-memory pipeline. Prefetching subsequent operand
tiles overlaps their transfer with computation on earlier stages.
Warps load packed fragments into registers and issue
\texttt{mma.sync} instructions with signed INT4 operands and
INT32 accumulation, avoiding explicit expansion of the operands
to BF16. 
At the instruction level, the dense \texttt{m16n8k64} MMA
form distributes a $16\times64$ activation fragment and a
$64\times8$ weight fragment across a 32-thread warp.
Each thread supplies four 32-bit registers containing packed
activation values and two containing packed weight values,
and holds four INT32 accumulators for the $16\times8$ output
fragment. Larger warp output tiles are assembled from
multiple such fragments. This mapping explains how packed
INT4 operands are consumed directly while accumulator
storage grows with the warp output-tile area.

Data-parallel execution completes each output tile within one
threadblock. Stream-K~\cite{streamk} may instead materialize partial accumulators
in a workspace when a tile spans multiple threadblocks.
After the required partial sums have been reconciled, the fused
epilogue applies the scales and stores the completed output
directly in BF16.

\begin{figure}[!t]
    \centering \includegraphics[width=1\linewidth]{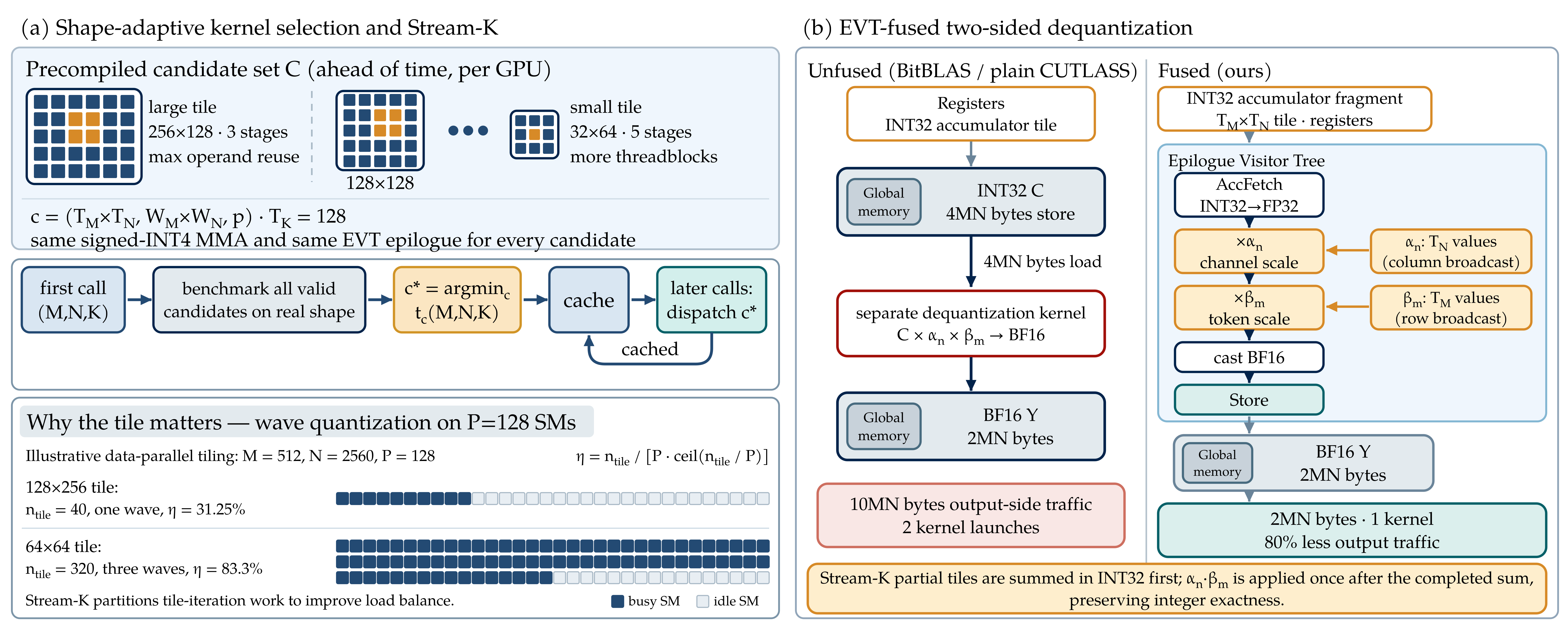}
    \vspace{-8mm}
    \caption{\footnotesize \textbf{Shape-adaptive execution and fused dequantization}.
(a) The lowest-latency valid configuration is selected and cached
for each shape and device. (b) EVT applies token and channel scales
in the GEMM epilogue and directly emits BF16, avoiding an intermediate
INT32 output tensor and a separate dequantization kernel.}
    \label{fig:sigprocess-i}
    \vspace{-3mm}
\end{figure}

\vspace{-1mm}
\subsection{Resource- and Shape-Adaptive Kernel Selection}
\label{sec:autotuning}

The fastest kernel configuration depends on both the GEMM shape
$(M,N,K)$ and the target GPU. Figure~\ref{fig:sigprocess-i}(a)
summarizes our selection flow: generate a set of precompiled
candidate kernels, discard candidates that violate hardware resource
or alignment constraints, benchmark the remaining candidates, and cache
the lowest-latency configuration for each shape and device. We represent a candidate as
$c=(T_M,T_N,T_K,W_M,W_N,p,d)$, where $T_M\times T_N$ and
$W_M\times W_N$ are the threadblock and warp output tiles,
respectively; $T_K$ is the reduction depth per mainloop iteration;
$p$ is the compile-time CUTLASS software-pipeline stage count; and
$d\in\{\mathrm{data\mbox{-}parallel},\mathrm{Stream\mbox{-}K}\}$
is the work-decomposition strategy. The stage count specifies the
number of shared-memory operand buffers used to overlap
\texttt{cp.async} transfers with MMA execution; it is not the physical
Tensor-Core pipeline depth. For $p$ stages, the nominal shared-memory payload occupied by packed
INT4 operands is $M_{\mathrm{pipe}}(c)=\frac{pT_K(T_M+T_N)}{2}
\quad\mathrm{bytes},$ excluding layout padding and any additional epilogue or synchronization
storage. 
Increasing $p$ adds buffering for latency hiding, whereas increasing
$T_K$ enlarges each operand tile and reduces the number of mainloop
iterations; both increase shared-memory consumption.
Register pressure is governed
mainly by the accumulator tile: a warp output tile of size
$W_M\times W_N$ contains $W_MW_N$ INT32 values distributed across
32 threads.
Thus, even though INT4 operands reduce operand-buffer storage relative
to BF16, INT32 accumulators can still limit occupancy.

Candidates that exceed the target GPU's register, shared-memory,
thread, or operand-alignment limits are excluded. For a valid candidate
$c$, we approximate the resident threadblocks per SM as
\begin{equation}
B_{\mathrm{res}}(c)=
\min\left(
\left\lfloor\frac{R_{\mathrm{SM}}}{R_c}\right\rfloor,
\left\lfloor\frac{M_{\mathrm{SM}}}{M_c}\right\rfloor,
\left\lfloor\frac{T_{\mathrm{SM}}}{T_c}\right\rfloor,
B_{\mathrm{SM}}^{\max}
\right),
\label{eq:resident_blocks}
\end{equation}
where $R_c$, $M_c$, and $T_c$ are the registers, shared-memory bytes,
and threads required by one threadblock, and
$R_{\mathrm{SM}}$, $M_{\mathrm{SM}}$, and $T_{\mathrm{SM}}$ are the
corresponding per-SM capacities. Because tile size, pipeline depth, and
occupancy interact non-monotonically, this model is used to filter
invalid candidates rather than to analytically choose the final kernel.

For data-parallel execution, the number of output tiles is
$n_{\mathrm{tile}}=\lceil M/T_M\rceil\lceil N/T_N\rceil$.
Assuming one resident threadblock per SM and equal tile execution
times, the idealized wave utilization on $P$ SMs is
$\eta=n_{\mathrm{tile}}/
[P\lceil n_{\mathrm{tile}}/P\rceil]$.
A partially occupied final wave reduces this utilization.
Stream-K can improve utilization by partitioning the finer-grained
tile-iteration work
$n_{\mathrm{work}}=n_{\mathrm{tile}}\lceil K/T_K\rceil$
across threadblocks, but it may introduce workspace traffic and
synchronization to reconcile partial INT32 accumulators. Its benefit
therefore depends on the full $(M,N,K)$ shape, not only on the number
of output tiles, so we evaluate both work-decomposition strategies when
they are valid.

During initialization or warm-up, the runtime benchmarks the valid
candidates and records
\begin{equation}
c^*(M,N,K,\mathcal{D})=
\arg\min_{c\in\mathcal{C}_{\mathcal{D}}}t_c(M,N,K),
\label{eq:best_configuration}
\end{equation}
where $\mathcal{D}$ identifies the target GPU and $t_c$ includes the
fused epilogue and any scheduling-related reduction overhead.
Subsequent invocations retrieve the cached configuration and dispatch
directly to $c^*$. 

\vspace{-1mm}
\subsection{EVT-Fused Two-Sided Dequantization}
\label{sec:evt}

As illustrated in Fig.~\ref{fig:sigprocess-i}(b), a conventional
implementation materializes the INT32 accumulator tensor and launches
a separate dequantization kernel, whereas EVT applies the scales in
the GEMM epilogue and directly emits BF16. Excluding scale-vector
loads and any Stream-K partial-reduction workspace traffic, the
output-side transfers are
\begin{align}
B_{\mathrm{unfused}}
&{=}(4_{\mathrm{store}}{+}4_{\mathrm{load}}{+}2_{\mathrm{BF16}})MN
{=}10MN~\mathrm{bytes}, \\
B_{\mathrm{fused}}&=2MN~\mathrm{bytes}.
\label{eq:epilogue_traffic}
\end{align}
Thus, EVT fusion eliminates $8MN$ bytes of intermediate traffic,
reducing these output-side transfers by $80\%$. We implement Eq.~\ref{eq:dequantization} through an Epilogue
Visitor Tree that converts the completed INT32 accumulator
to FP32, applies the channel and token scales, and casts the
result to BF16:
\begin{equation}
 A_{m,n}
 \rightarrow
 \mathrm{FP32}(A_{m,n})
 \times \alpha_n
 \times \beta_m
 \rightarrow
 \mathrm{BF16}(Y_{m,n}).
\end{equation}
The channel scale $\alpha_n$ is broadcast along the $M$ dimension and
the token scale $\beta_m$ is broadcast along the $N$ dimension.
Scaling is performed in FP32 on register-resident accumulator
fragments before the result is converted to BF16.

For an interior $T_M\times T_N$ output tile, the epilogue
requires $T_M$ distinct token scales and $T_N$ distinct channel
scales. Each token scale applies to $T_N$ outputs, while each
channel scale applies to $T_M$ outputs. The visitors associate
these scales with the logical coordinates of the accumulator
fragments, forming scale products as the fragments are
processed. This requires only the two scale vectors, without
materializing an $M\times N$ matrix of scale products. When Stream-K divides a tile across threadblocks, its
accumulator can be written as $A_{m,n}=\sum_{r\in\mathcal{R}_{m,n}}A^{(r)}_{m,n}$, where $\mathcal{R}_{m,n}$ indexes the contributing reduction
partitions. Since $\alpha_n\beta_m$ is independent of the
partition, scaling can follow the completed integer reduction.
Keeping partial sums in INT32 preserves the integer result
under the accumulator bound established earlier and avoids
rounding individual contributions to BF16 before summation.
The fused epilogue therefore eliminates the output buffer
used solely for dequantization while retaining any workspace
required for partial-sum reconciliation.


Because the EVT inherits the output-fragment
layout of its parent GEMM, the same dequantization expression is
instantiated for every candidate kernel configuration.

\subsection{End-to-End Operator Boundary}

The reported W1.58A4 prefill latency includes online activation-scale computation, activation quantization and packing, INT4 Tensor Core
GEMM, scale application, and BF16 output storage. Offline weight quantization, persistent weight packing, and
one-time kernel configuration search are excluded from
steady-state inference latency because their costs are
amortized across requests. Nonlinear operations, attention,
normalization, residual connections, embeddings, and the output head
retain their original model precision unless otherwise stated. 
This boundary distinguishes operator latency from end-to-end model
latency and supports analysis of quantization, GEMM, and epilogue costs.

\section{Experiment}
 We evaluate the proposed method at three levels: the INT4 GEMM kernel in isolation, the full BitNet-b1.58 forward pass (prefill / decode), and ablation study on the proposed shape-adaptive kernel selection and EVT-fused dequantization.

\vspace{-1mm}
\subsection{Experimental Setup}

\begin{table}[!t]
\caption{Kernel Level Comparison. All kernels are evaluated with the shape $(M, N, K)$ while $N$ and $K$ are fixed to 2560. Experiments are conducted on a single RTX 4090.}
\label{tab: kernel level comparison}
\vspace{-4mm}
\begin{center}
\setlength{\tabcolsep}{1mm}{
\footnotesize
\begin{tabular}{l|c|c|c|c|c}
\hline
\hline

 $M$ & Kernel & Quant+Pack ($\mu$s) & GEMM ($\mu$s) & Dequant ($\mu$s) & Total ($\mu$s) \\
\hline
 \multirow{3}{*}{64} & BF16 & - & 12.8 & - & 12.8 \\
 & BitBLAS & 3.0 & 6.3 & 9.5 & 18.8 \\
 \cline{4-5}
 & Ours & 4.2 & \multicolumn{2}{c|}{7.6} & \textbf{11.8} \\

\hline

 \multirow{3}{*}{512} & BF16 & - & 57.8 & - & 57.8 \\
 & BitBLAS & 5.6 & 17.1 & 20.2 & 42.9 \\
 \cline{4-5}
 & Ours & 7.8 & \multicolumn{2}{c|}{18.8} & \textbf{26.6} \\

\hline

 \multirow{3}{*}{1024} & BF16 & - & 95.6 & - & 95.6 \\
 & BitBLAS & 8.0 & 24.5 & 30.8 & 63.3 \\
 \cline{4-5}
 & Ours & 9.8 & \multicolumn{2}{c|}{29.9} & \textbf{39.7} \\

\hline

 \multirow{3}{*}{2048} & BF16 & - & 183.0 & - & 183.0 \\
 & BitBLAS & 27.9 & 49.4 & 61.4 & 138.7 \\
 \cline{4-5}
 & Ours & 15.6 & \multicolumn{2}{c|}{44.5} & \textbf{60.1} \\

\hline
\hline
\end{tabular}
}
\end{center}
\vspace{-8mm}
\end{table}

\noindent\textbf{Hardware}: Experiments are conducted on an NVIDIA RTX 4090 (Ada Lovelace, 128 SMs, 24\,GB GDDR6X, 1008\,GB/s) and a Jetson Orin Nano 8GB (Ampere, 8 SMs, 8\,GB LPDDR5, 68\,GB/s), both supporting native INT4 tensor-core MMA. These platforms represent distinct execution regimes: the RTX 4090's high parallelism makes incomplete threadblock waves important for small and irregular GEMMs, whereas the Jetson's limited SM count, shared memory, and power budget emphasize tile resource usage and occupancy. We therefore optimize kernel configurations independently for each device.
 We use CUDA 12.6, CUTLASS 4.6.0, with JetPack 6.2.2 on Jetson.

\noindent\textbf{Model}: We evaluate the proposed kernel on BitNet-b1.58~\cite{bitnet1_58} model.
On RTX 4090, we evaluate the original 3B model with all kernels.
On Jetson Orin Nano with 8\,GB DRAM, we evaluate the 3B model
with our method and the 700M model across the runnable baselines,
as the evaluated BF16 and BitNet implementations exceed the
available memory when running the 3B model.

\noindent\textbf{Supervised Fine-Tuning}: We perform quantization-aware SFT on pretrained ternary BitNet b1.58 models (700M and 3B), following the 1bitLLM reproduction~\cite{bitnet1_58}, with activations further quantized to 4 bits (A4).
The 85,805-example corpus combines instruction-following, multi-turn dialogue, and reasoning data from ShareGPT~\cite{vicuna2023}, Alpaca~\cite{taori2023alpaca}, GSM8K~\cite{cobbe2021gsm8k}, Anthropic HH~\cite{bai2022helpful}, and OpenAssistant~\cite{kopf2023openassistant}. Training uses a maximum length of 2048 tokens, 3 epochs, AdamW, a peak learning rate of $1\times10^{-4}$, cosine scheduling, and a 0.1 warmup ratio. 

\noindent\textbf{Baselines}: We compare the proposed kernel with the following: i) \textbf{BF16}: cuBLAS bf16 kernel for both prefill and decode phases, ii) \textbf{BitNet}~\cite{bitnet1_58}: the official BitNet-b1.58 implementation, which deploys W1.58A8 `dp4a' kernel for decode phase while bf16 kernel for prefill phase, and iii) \textbf{BitBLAS}~\cite{bitblas}: BitBLAS dynamically generated kernels for prefill.
 
\noindent\textbf{Metrics}: We report four metrics: (1) \textit{Latency}: Kernel latency is averaged over 50 timed launches after warm-up; model-level prefill and decode latencies are averaged over three warmed-up runs of the full generation pipeline with batch size $1$ and $32$ generated tokens. (2) \textit{GPU power}: Mean power is measured by replaying each phase's captured CUDA graph during 3\,s, sampled every 20\,ms (NVML board power). (3) \textit{Energy}: We report energy per prefill and generated token, using NVML's cumulative energy counter on RTX 4090 and mean rail power multiplied by per-call time on Jetson, which lacks an energy counter. (4) \textit{Perplexity (PPL)}: WikiText-2 \cite{wikitext} PPL measures quantized-model accuracy relative to the baselines under each configuration.

\begin{table*}[!h]
\caption{LLM Full Benchmark. All methods are measured for a single round response with prompt length $512$. `()' shows the speedup v.s. BF16 baseline.}
\label{tab: llm benchmark}
\vspace{-5mm}
\begin{center}
\setlength{\tabcolsep}{2mm}{
\footnotesize
\begin{tabular}{l|c|c|c|c|c|c|c|c|c}
\hline
\hline

 \multirow{2}{*}{Device} & \multirow{2}{*}{Model} & \multirow{2}{*}{Kernel} & \multicolumn{2}{c|}{Latency (ms) $\downarrow$} & \multicolumn{2}{c|}{GPU Power (W) $\downarrow$} & \multicolumn{2}{c|}{Energy (J) $\downarrow$} & \multirow{2}{*}{PPL$\downarrow$} \\
 \cline{4-9}
 & & & Prefill & Decode & Prefill & Decode & Prefill & Decode & \\

\hline

 \multirow{4}{*}{RTX 4090} & \multirow{4}{*}{BitNetb1.58-3B} & BF16 & 41.10 & 11.79 & 350.44 & 268.97 & 17.12 & 2.97 & 10.04 \\
 & & BitNet~\cite{bitnet1_58} & 41.14 & 5.73 & 337.86 & 215.53 & 17.55 & \textbf{1.02} & \textbf{9.91} \\
 & & BitBLAS~\cite{bitblas} & 36.34 & 11.81 & 346.59 & 264.89 & 13.21 & 3.00 & 9.97 \\
 & & Ours & \textbf{28.94} ($1.42\times$) & \textbf{5.73} ($2.06\times$) & \textbf{274.64} & \textbf{202.18} & \textbf{9.99} & 1.04 & 10.05  \\

\hline

 \multirow{8}{*}{Jetson Orin Nano} & \multirow{4}{*}{BitNetb1.58-3B} & BF16 & \multicolumn{7}{c}{Out of Memory} \\
 & & BitNet~\cite{bitnet1_58} & \multicolumn{7}{c}{Out of Memory} \\
 & & BitBLAS~\cite{bitblas} & \multicolumn{7}{c}{Unsupported on Jetson platform} \\
 \cline{4-10}
 & & Ours & 276.69 & 33.98 & 6.58 & 6.87 & 1.82 & 0.23 & 10.05 \\

\cline{2-10}

 & \multirow{4}{*}{BitNetb1.58-700M} & BF16 & 120.52 & 18.52 & 9.33 & 7.89 & 1.12 & 0.15 & \textbf{12.33} \\
 & & BitNet~\cite{bitnet1_58} & 117.57 & 7.45 & 8.98 & 7.51 & 1.05 & 0.06 & 12.87 \\
 \cline{4-10}
 & & BitBLAS~\cite{bitblas} & \multicolumn{7}{c}{Unsupported on Jetson platform} \\
 \cline{4-10}
 & & Ours & \textbf{106.24} ($1.13\times$) & \textbf{7.34} ($2.52\times$) & \textbf{6.06} & \textbf{6.83} & \textbf{0.64} & \textbf{0.05} & 12.66 \\

\hline
\hline
\end{tabular}
}
\end{center}
\vspace{-8mm}
\end{table*}

\vspace{-1mm}
\subsection{Kernel-Level Comparison}
 We firstly conduct kernel-level comparison for INT4xINT4 GEMM-kernel, i.e., BitBLAS and ours kernel, v.s. BF16 cuBLAS kernel.
 As a strongest baseline, BitBLAS runs the same INT4xINT4 matrix multiply-accumulate on the same packed operands.
 The core difference between BitBLAS kernel and our method is that BitBLAS is a general mixed-precision library, whose ladder type system and TileLang code generator simultaneously serve FP16xINT4, INT8xINT2 weight-only GEMMs, dequantizing GEMVs for decode, and native INT4xINT4 GEMM kernel.
 All are derived from one generic matmul template whose tiles and pipeline are proposed by a hardware-cost-model heuristic and then ranked by latency, while the dequantization is built around weight dequantization (group scales, zero points) instead of activation dequantization.
 Our kernel is optimized for W1.58A4 prefill, and two mechanisms of Sec.~\ref{sec:method} follow from it.
 We breakdown each kernel into three phases: `Quant+Pack', `GEMM INT4xINT4 Matrix-Multiply Accumulate', and `Dequant'.
 For BF16, there is no quantization/pack and dequantization operations, while our method fuse dequantization into GEMM.
 As shown in Table~\ref{tab: kernel level comparison}, our dequantization-fused GEMM kernel achieves about $2\times$ acceleration compared to BitBLAS kernel.





\vspace{-1mm}
\subsection{LLM Benchmark.}
 We run the full LLM model with a 512-token prompt and report latency, GPU power, energy for prefill and decode phase, respectively.
 We also report perplexity (PPL) as the accuracy benchmark of the model integrated with each kernel.
 For the W1.58A4 configuration, we follow the BitNet quantization-aware training (QAT) scheme and quantize activations to 4 bits during SFT. 
 Specifically, activations are dynamically quantized using a per-token absmax scaling factor, while a straight-through estimator (STE)\cite{ste} is applied through the quantization operation during backpropagation in SFT process. 
 Perplexities are evaluated without additional post-training quantization or calibration is applied after SFT. 
 All methods are measured on a RTX4090 and Jetson Orin Nano, respectively.
 Additionally, the Jetson Orin Nano has only 8GB unified memory, which cannot hold the half-precision (BF16) BitNet1.58b-3B and the BitNet (W1.58A8), while there is no supported BitBLAS kernel on Jetson.
 Our method is the only configuration that can deploy the BitNet1.58b-3B model on Jetson.
 To compare with other baselines on Jetson, we also report a smaller model BitNet1.58-700M with reduced number of layers and vocabulary size.
As shown in Table~\ref{tab: llm benchmark}, our method reduces
3B-model prefill latency on RTX 4090 from 41.10 to 28.94\,ms
($1.42\times$), while lowering prefill GPU power and energy
relative to all evaluated baselines. Decode retains the BitNet
kernel and matches its 5.73\,ms latency. On Jetson Orin Nano, the 700M model achieves
a $1.13\times$ prefill speedup over BF16.

\vspace{-2mm}
\subsection{Ablation Study}

\noindent\textbf{Module Discussion}: We conduct the ablation study to investigate the improvement brought by our proposed Shape-Adaptive Kernel Selection (SAKS) and Epilogue Visitor Tree fused dequantization (EVT-deq).
 As shown in Table~\ref{tab: ablation study}, SAKS reduces the latency from 33.59 ms to 29.35 ms, Stream-K reduces the latency from 33.59 ms to 32.13 ms, and reduce the GPU power from 324.10 W to 275.71 W.
 With applying EVT-fused dequantization, the latency is reduced from 33.59 ms to 31.43 ms.
 By integrating all these two modules, our method finally achieves the lowest latency and GPU power.

\begin{table}[!t]
\caption{\footnotesize Module Discussion. The latency and GPU Power are reported for prefill phase with prompt length 512 of BitNetb1.58-3B model running on RTX 4090.}
\label{tab: ablation study}
\vspace{-4mm}
\begin{center}
\setlength{\tabcolsep}{1mm}{
\footnotesize
\begin{tabular}{c|c|c|c|c|c}
\hline
\hline

GEMM & SAKS & Stream-K & EVT-deq & Latency (ms)$\downarrow$ & GPU Power (W) $\downarrow$ \\

\hline

$\checkmark$ & $\times$ & $\times$ & $\times$ & 33.59 & 324.10 \\
$\checkmark$ & $\checkmark$ & $\times$ & $\times$ & 29.35 & 275.71 \\
$\checkmark$ & $\times$ & $\checkmark$ & $\times$ & 32.13 & 282.82 \\
$\checkmark$ & $\times$ & $\times$ & $\checkmark$ & 31.43 & 344.00 \\
$\checkmark$ & $\checkmark$ & $\checkmark$ & $\checkmark$ & \textbf{28.94} & \textbf{274.64} \\

\hline
\hline
\end{tabular}
}
\end{center}
\vspace{-6mm}
\end{table}

\begin{figure}
    \centering \includegraphics[width=0.9\linewidth]{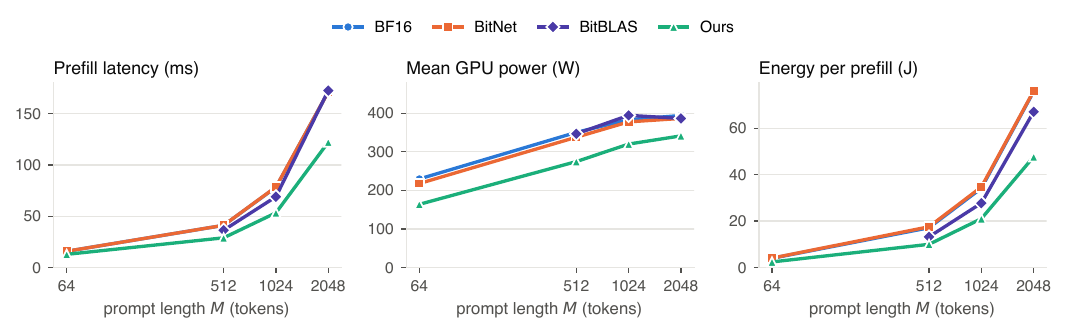}
    \vspace{-3mm}
    \caption{\footnotesize Prompt Length Sweeping.
    Latency, GPU Power, and Energy are reported for prefill phase of BitNetb1.58-3B model running on RTX 4090.
    Ours method achieves the best performance over latency, GPU power, and Energy consumption.}
    \label{fig:prompt_length_sweeping}
    \vspace{-4mm}
\end{figure}

\noindent\textbf{Prompt Length Discussion}: As discussed in Sec.~\ref{sec:arithmetic-intensity motivation}, increasing $M$ would amortize weights read across more tokens and increase arithmetic intensity.
 Ideally, prefill will benefit from our INT4xINT4 GEMM kernel much more with $M$ increasing.
 We conduct such ablation study and report different $M$ settings, i.e., prompt length, for each kernel method.
 As shown in Fig.~\ref{fig:prompt_length_sweeping}, our method achieves the best performance for various $M$ from $64$ to $2048$.

\begin{figure}[!t]
    \centering \includegraphics[width=0.9\linewidth]{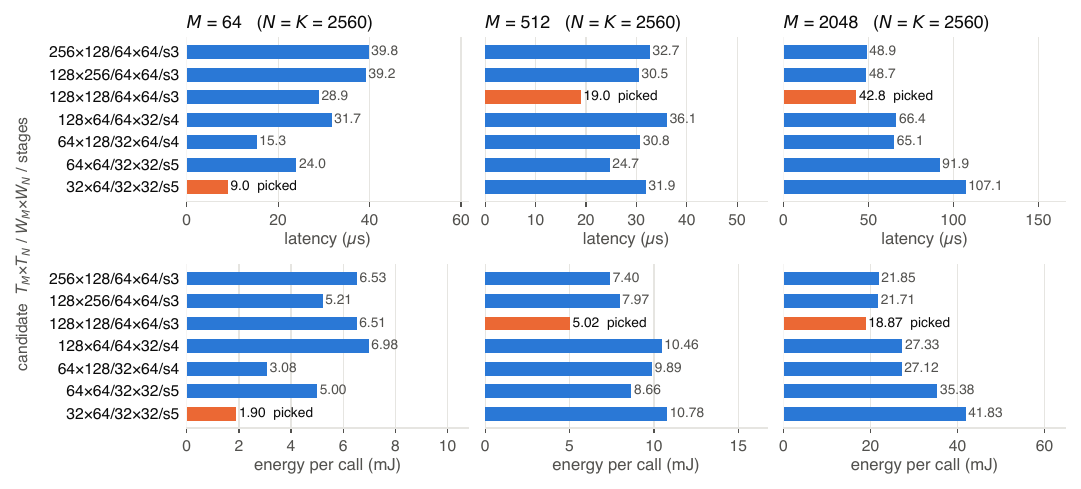}
    \vspace{-5mm}
    \caption{\footnotesize Tile-size hit analysis for SAKS. Our proposed SAKS module could pick the optimal candidate tile size in various kernel shape settings.}
    \label{fig:tile_sweeping}
    \vspace{-5mm}
\end{figure}

 \noindent\textbf{Tile-size Hit Analysis}: We analyze the latency and energy consumption for different candidate tile-sizes.
 As shown in Fig.~\ref{fig:tile_sweeping}, our proposed Shape-Adaptive Kernel Selection (SAKS) hits the optimal tile size for various kernel shapes.

\vspace{-1mm}
\section{Conclusion}

We presented a W1.58A4 operator that moves the compute-bound prefill of ternary LLMs onto native INT4 tensor cores, where weight-only quantization brings no gain. Shape-adaptive kernel selection with Stream-K keeps the tensor cores occupied across prefill lengths, and an EVT-fused epilogue applies the per-token and per-channel scales in registers, removing the separate dequantization pass that limits BitBLAS at long prompts. On an RTX 4090 the operator is about $2\times$ faster than BitBLAS and accelerates BitNet-b1.58-3B prefill by $1.42\times$ over BF16 at $42\%$ less energy, with similar perplexity; on a Jetson Orin Nano it is the only configuration that fits the 3B model in 8~GB. 

\bibliographystyle{IEEEtran}
\bibliography{ref}

\begin{thebibliography}{10}
\providecommand{\url}[1]{#1}
\csname url@samestyle\endcsname
\providecommand{\newblock}{\relax}
\providecommand{\bibinfo}[2]{#2}
\providecommand{\BIBentrySTDinterwordspacing}{\spaceskip=0pt\relax}
\providecommand{\BIBentryALTinterwordstretchfactor}{4}
\providecommand{\BIBentryALTinterwordspacing}{\spaceskip=\fontdimen2\font plus
\BIBentryALTinterwordstretchfactor\fontdimen3\font minus \fontdimen4\font\relax}
\providecommand{\BIBforeignlanguage}[2]{{%
\expandafter\ifx\csname l@#1\endcsname\relax
\typeout{** WARNING: IEEEtran.bst: No hyphenation pattern has been}%
\typeout{** loaded for the language `#1'. Using the pattern for}%
\typeout{** the default language instead.}%
\else
\language=\csname l@#1\endcsname
\fi
#2}}
\providecommand{\BIBdecl}{\relax}
\BIBdecl

\bibitem{bitnetb1_58-2b4t}
\BIBentryALTinterwordspacing
S.~Ma, H.~Wang, S.~Huang, X.~Zhang, Y.~Hu, T.~Song, Y.~Xia, and F.~Wei, ``Bitnet b1.58 2b4t technical report,'' 2025. [Online]. Available: \url{https://arxiv.org/abs/2504.12285}
\BIBentrySTDinterwordspacing

\bibitem{bitnet_cpp}
J.~Wang, H.~Zhou, T.~Song, S.~Cao, Y.~Xia, T.~Cao, J.~Wei, S.~Ma, H.~Wang, and F.~Wei, ``Bitnet. cpp: Efficient edge inference for ternary llms,'' in \emph{Proceedings of the 63rd Annual Meeting of the Association for Computational Linguistics (Volume 1: Long Papers)}, 2025, pp. 9305--9322.

\bibitem{bitnet1_58}
\BIBentryALTinterwordspacing
S.~Ma, H.~Wang, L.~Ma, L.~Wang, W.~Wang, S.~Huang, L.~Dong, R.~Wang, J.~Xue, and F.~Wei, ``The era of 1-bit llms: All large language models are in 1.58 bits,'' 2024. [Online]. Available: \url{https://arxiv.org/abs/2402.17764}
\BIBentrySTDinterwordspacing

\bibitem{onebit}
Y.~Xu, X.~Han, Z.~Yang, S.~Wang, Q.~Zhu, Z.~Liu, W.~Liu, and W.~Che, ``Onebit: Towards extremely low-bit large language models,'' \emph{Advances in Neural Information Processing Systems}, vol.~37, pp. 66\,357--66\,382, 2024.

\bibitem{pbllm}
Z.~Yuan, Y.~Shang, and Z.~Dong, ``Pb-llm: Partially binarized large language models,'' in \emph{The Twelfth International Conference on Learning Representations}, 2024.

\bibitem{billm}
\BIBentryALTinterwordspacing
W.~Huang, Y.~Liu, H.~Qin, Y.~Li, S.~Zhang, X.~Liu, M.~Magno, and X.~Qi, ``Billm: Pushing the limit of post-training quantization for llms,'' 2024. [Online]. Available: \url{https://arxiv.org/abs/2402.04291}
\BIBentrySTDinterwordspacing

\bibitem{zhang2025systematic}
H.~Zhang, P.~S. Yu, and J.~Zhang, ``A systematic survey of text summarization: From statistical methods to large language models,'' \emph{ACM Computing Surveys}, vol.~57, no.~11, pp. 1--41, 2025.

\bibitem{zhang2024benchmarking}
T.~Zhang, F.~Ladhak, E.~Durmus, P.~Liang, K.~McKeown, and T.~B. Hashimoto, ``Benchmarking large language models for news summarization,'' \emph{Transactions of the Association for Computational Linguistics}, vol.~12, pp. 39--57, 2024.

\bibitem{bai2025longbench}
Y.~Bai, S.~Tu, J.~Zhang, H.~Peng, X.~Wang, X.~Lv, S.~Cao, J.~Xu, L.~Hou, Y.~Dong \emph{et~al.}, ``Longbench v2: Towards deeper understanding and reasoning on realistic long-context multitasks,'' in \emph{Proceedings of the 63rd Annual Meeting of the Association for Computational Linguistics (Volume 1: Long Papers)}, 2025, pp. 3639--3664.

\bibitem{ling2025longreason}
Z.~Ling, K.~Liu, K.~Yan, Y.~Yang, W.~Lin, T.-H. Fan, L.~Shen, Z.~Du, and J.~Chen, ``Longreason: A synthetic long-context reasoning benchmark via context expansion,'' \emph{arXiv preprint arXiv:2501.15089}, 2025.

\bibitem{wu2023next}
S.~Wu, H.~Fei, L.~Qu, W.~Ji, and T.-S. Chua, ``Next-gpt: Any-to-any multimodal llm,'' \emph{arXiv preprint arXiv:2309.05519}, 2023.

\bibitem{zhan2024anygpt}
J.~Zhan, J.~Dai, J.~Ye, Y.~Zhou, D.~Zhang, Z.~Liu, X.~Zhang, R.~Yuan, G.~Zhang, L.~Li \emph{et~al.}, ``Anygpt: Unified multimodal llm with discrete sequence modeling,'' in \emph{Proceedings of the 62nd Annual Meeting of the Association for Computational Linguistics (Volume 1: Long Papers)}, 2024, pp. 9637--9662.

\bibitem{islam2025gpt}
R.~Islam and O.~M. Moushi, ``Gpt-4o: The cutting-edge advancement in multimodal llm,'' in \emph{Intelligent Computing-Proceedings of the Computing Conference}.\hskip 1em plus 0.5em minus 0.4em\relax Springer, 2025, pp. 47--60.

\bibitem{streamk}
M.~Osama, D.~Merrill, C.~Cecka, M.~Garland, and J.~D. Owens, ``Stream-k: Work-centric parallel decomposition for dense matrix-matrix multiplication on the gpu,'' in \emph{Proceedings of the 28th ACM SIGPLAN Annual Symposium on Principles and Practice of Parallel Programming}, 2023, pp. 429--431.

\bibitem{bitnet}
\BIBentryALTinterwordspacing
H.~Wang, S.~Ma, L.~Dong, S.~Huang, H.~Wang, L.~Ma, F.~Yang, R.~Wang, Y.~Wu, and F.~Wei, ``Bitnet: Scaling 1-bit transformers for large language models,'' 2023. [Online]. Available: \url{https://arxiv.org/abs/2310.11453}
\BIBentrySTDinterwordspacing

\bibitem{spectra}
\BIBentryALTinterwordspacing
A.~Kaushal, T.~Vaidhya, A.~K. Mondal, T.~Pandey, A.~Bhagat, and I.~Rish, ``Spectra: Surprising effectiveness of pretraining ternary language models at scale,'' 2024. [Online]. Available: \url{https://arxiv.org/abs/2407.12327}
\BIBentrySTDinterwordspacing

\bibitem{paretoq}
Z.~Liu, C.~Zhao, H.~Huang, S.~Chen, J.~Zhang, J.~Zhao, S.~Roy, L.~Jin, Y.~Xiong, Y.~Shi \emph{et~al.}, ``Paretoq: Improving scaling laws in extremely low-bit llm quantization,'' \emph{Advances in Neural Information Processing Systems}, vol.~38, pp. 91\,311--91\,336, 2026.

\bibitem{bitnet_a4_8}
\BIBentryALTinterwordspacing
H.~Wang, S.~Ma, and F.~Wei, ``Bitnet a4.8: 4-bit activations for 1-bit llms,'' 2024. [Online]. Available: \url{https://arxiv.org/abs/2411.04965}
\BIBentrySTDinterwordspacing

\bibitem{tmac}
J.~Wei, S.~Cao, T.~Cao, L.~Ma, L.~Wang, Y.~Zhang, and M.~Yang, ``T-mac: Cpu renaissance via table lookup for low-bit llm deployment on edge,'' in \emph{Proceedings of the Twentieth European Conference on Computer Systems}, 2025, pp. 278--292.

\bibitem{luttensorcore}
Z.~Mo, L.~Wang, J.~Wei, Z.~Zeng, S.~Cao, L.~Ma, N.~Jing, T.~Cao, J.~Xue, F.~Yang \emph{et~al.}, ``Lut tensor core: A software-hardware co-design for lut-based low-bit llm inference,'' in \emph{Proceedings of the 52nd Annual International Symposium on Computer Architecture}, 2025, pp. 514--528.

\bibitem{cutlass}
\BIBentryALTinterwordspacing
V.~Thakkar, P.~Ramani, C.~Cecka, A.~Shivam, H.~Lu, and E.~Yan, ``Cutlass.'' [Online]. Available: \url{https://github.com/NVIDIA/cutlass}
\BIBentrySTDinterwordspacing

\bibitem{bitblas}
L.~Wang, L.~Ma, S.~Cao, Q.~Zhang, J.~Xue, Y.~Shi, N.~Zheng, Z.~Miao, F.~Yang, T.~Cao \emph{et~al.}, ``Ladder: Enabling efficient low-precision deep learning computing through hardware-aware tensor transformation,'' in \emph{18th USENIX Symposium on Operating Systems Design and Implementation (OSDI 24)}, 2024, pp. 307--323.

\bibitem{evt}
Z.~Chen, A.~Kerr, R.~Cai, J.~Kosaian, H.~Wu, Y.~Ding, and Y.~Xie, ``Evt: Accelerating deep learning training with epilogue visitor tree,'' in \emph{Proceedings of the 29th ACM International Conference on Architectural Support for Programming Languages and Operating Systems, Volume 3}, 2024, pp. 301--316.

\bibitem{vicuna2023}
W.-L. Chiang, Z.~Li, Z.~Lin, Y.~Sheng, Z.~Wu, H.~Zhang, L.~Zheng, S.~Zhuang, Y.~Zhuang, J.~E. Gonzalez, I.~Stoica, and E.~P. Xing, ``Vicuna: An open-source chatbot impressing gpt-4 with 90\%* chatgpt quality,'' \url{https://lmsys.org/blog/2023-03-30-vicuna/}, 2023.

\bibitem{taori2023alpaca}
R.~Taori, I.~Gulrajani, T.~Zhang, Y.~Dubois, X.~Li, C.~Guestrin, P.~Liang, and T.~B. Hashimoto, ``Alpaca: A strong, replicable instruction-following model,'' \url{https://crfm.stanford.edu/2023/03/13/alpaca.html}, 2023.

\bibitem{cobbe2021gsm8k}
K.~Cobbe, V.~Kosaraju, M.~Bavarian, M.~Chen, H.~Jun, L.~Kaiser, M.~Plappert, J.~Tworek, J.~Hilton, R.~Nakano, C.~Hesse, and J.~Schulman, ``Training verifiers to solve math word problems,'' \emph{arXiv preprint arXiv:2110.14168}, 2021.

\bibitem{bai2022helpful}
Y.~Bai, A.~Jones, K.~Ndousse, A.~Askell, A.~Chen, N.~DasSarma, D.~Drain, D.~Ganguli, T.~Henighan, N.~Joseph, S.~Kadavath \emph{et~al.}, ``Training a helpful and harmless assistant with reinforcement learning from human feedback,'' \emph{arXiv preprint arXiv:2204.05862}, 2022.

\bibitem{kopf2023openassistant}
A.~K{\"o}pf, Y.~Kilcher, D.~von R{\"u}tte, S.~Anagnostidis, Z.-R. Tam, K.~Stevens, A.~Barhoum, N.~M. Duc, O.~Stanley, R.~Nagyfi \emph{et~al.}, ``Openassistant conversations---democratizing large language model alignment,'' in \emph{Advances in Neural Information Processing Systems}, 2023.

\bibitem{wikitext}
S.~Merity, C.~Xiong, J.~Bradbury, and R.~Socher, ``Pointer sentinel mixture models,'' in \emph{International Conference on Learning Representations (ICLR)}, 2017.

\bibitem{ste}
Y.~Bengio, N.~L{'e}onard, and A.~Courville, ``Estimating or propagating gradients through stochastic neurons for conditional computation,'' \emph{arXiv preprint arXiv:1308.3432}, 2013.

\end{thebibliography}

\end{document}